\documentclass[10pt, twocolumn, comsoc]{IEEEtran}

\usepackage{graphicx,epsfig}
\usepackage[noadjust]{cite}
\usepackage{mcite}
\usepackage{amsfonts,helvet}
\usepackage{fancyhdr}
\usepackage{threeparttable}
\usepackage{epsf,epsfig}
\usepackage{amsthm}
\usepackage{amsmath}
\usepackage{siunitx}
\usepackage{amssymb}
\usepackage{stfloats}

\usepackage{dsfont}
\usepackage{subfigure}
\usepackage{color}
\usepackage{enumerate}
\usepackage{gensymb}
\usepackage{cancel}
\usepackage{lipsum}
\usepackage{mathtools}
\usepackage{cuted}
\usepackage{bbm}
\usepackage[linesnumbered,ruled]{algorithm2e}

\usepackage{eucal}

\begin{document}

\title{Uplink Coordinated Pilot Design for 1-bit Massive MIMO in Correlated Channel}

\author{Hyeongtak~Yun, Juntaek~Han, Kaiming~Shen, and Jeonghun~Park

\thanks{
This work was supported in part by Institute of Information \& communications Technology Planning \& Evaluation (IITP) grant funded by the Korea government (MSIT) (No. RS-2024-00397216, Development of the Upper-mid Band Extreme massive MIMO (E-MIMO), No. 2025-RS-2024-00428780, 6G Cloud Research and Education Open Hub and No. RS-2024-00434743, YKCS Open RAN Global Collaboration Center).
H. Yun, J. Han, and J. Park are with School of Electrical and Electronic Engineering, Yonsei University, Seoul, South Korea (e-mail:{\texttt{ yht3114, jthan1218, jhpark@yonsei.ac.kr}}). {\color{black}{K. Shen is with School of Science and Engineering, The Chinese University of Hong Kong, Shenzhen, China (e-mail: {\texttt{shenkaiming@cuhk.edu.cn}}).}}
}
}
 
\maketitle \setcounter{page}{1} 
\begin{abstract} 
In this paper, we propose a coordinated pilot design method to minimize the channel estimation mean squared error (MSE) in 1-bit analog-to-digital converters (ADCs) massive multiple-input multiple-output (MIMO).
Under the assumption that the well-known Bussgang linear minimum mean square error (BLMMSE) estimator is used for channel estimation, we first observe that the resulting MSE leads to an intractable optimization problem, as it involves the arcsin function and a complex multiple matrix ratio form. 
To resolve this, we derive the approximate MSE by assuming the low signal-to-noise ratio (SNR) regime, by which we develop an efficient coordinated pilot design based on a fractional programming technique. 
The proposed pilot design is distinguishable from the existing work in that it is applicable in general system environments, including correlated channel and multi-cell environments. 
We demonstrate that the proposed method outperforms the channel estimation accuracy performance compared to the conventional approaches. 
\end{abstract}

\section{Introduction}
In multi-user multiple-input multiple-output (MU-MIMO) systems, channel state information (CSI) acquisition is a critical part to achieve high spectral efficiency performance, as imperfect CSI incurs additional interference, which severely limits the achievable spectral efficiency \cite{park:twc:16, caire:tit:10}.
In cellular networks, CSI estimation performance is mainly determined by two key factors: i) the channel estimation algorithm and ii) the pilot sequence. 
If the channel vector is modeled as a Gaussian distribution, it has been well studied that the optimal mean squared error (MSE) performance is achievable by using the linear minimum mean squared error (LMMSE) estimator \cite{yin:jsac:2013}. This paper focuses on the pilot sequence aspect.

Typically, if sufficient time and frequency resources are available compared to the number of users, assigning mutually orthogonal pilots to all the users is optimal in terms of the MSE.  
However, when the pilots are reused in multi-cell environments \cite{marzetta:twc:10}, i.e., pilot contamination occurs, 
it is infeasible to ensure orthogonality between the pilot sequences. 
This non-orthogonality leads to significant interference during channel estimation, resulting in a notable increase in MSE.
In such cases, it becomes crucial to design pilot sequences in a coordinated manner taking into account both channel estimation accuracy and mitigation of pilot interference. This motivates several advanced pilot design methods.
In particular, it has been shown that coordinated pilot design across multiple cells to minimize overall MSE is effective in mitigating pilot contamination by minimizing the overall MSE \cite{al-salihi:commlett:18, stein:tcom:20, kaiming_coordinated, wu:icassp:20, lim:tvt:22}.   
Notably, in \cite{kaiming_coordinated}, a fractional programming (FP) based pilot design technique via a matrix quadratic transform technique \cite{kaiming:tsp:fp} was devised for coordinated pilot sequence optimization.
A common consideration of the above mentioned prior work is ideal hardware MIMO systems, such as an analog-to-digital converter (ADC) with no quantization error. 

Recently, massive MIMO with low resolution ADCs (particularly 1-bit ADCs) has gained significant attention thanks to its energy and hardware efficiency \cite{walden:jsac:99, choi:twc:22, mezghani:isit:09}.
When the channel estimation is performed on the basis of 1-bit quantized measurements, the nonlinear nature of quantization makes the observations no longer follow a Gaussian distribution, so as to render conventional LMMSE channel estimation far from optimal.
Addressing this, one promising approach is the Bussgang LMMSE (BLMMSE) \cite{yongzhi:tsp:17, wan:twc:20}. 
The key idea of the BLMMSE is to use the Bussgang decomposition, wherein the output of the nonlinear function can be expressed as the sum of linear terms and an uncorrelated distortion term, enabling the derivation of LMMSE estimators for systems with nonlinearities.

One essential aspect in BLMMSE for 1-bit ADC channel estimation is that the MSE characterization is significantly changed. To be specific, the analytical form of the MSE achieved by BLMMSE is determined by the normalized autocorrelation matrix of the output signal, which is characterized by an element-wise arcsin function \cite{yongzhi:tsp:17}.
As a result, the existing pilot design approaches, which are based on ideal hardware assumptions such as high-resolution ADCs, are not directly applicable.
To resolve this issue, \cite{yi_one_bit_Rayleigh} developed a pilot design method that accounts for the non-linearity of 1-bit ADCs and minimizes the MSE under a hypersphere constraint via a steepest-descent approach. 
Despite this, it is only applicable to limited environments, such as uncorrelated Rayleigh channel and a single cell environment. By considering that the pilot contamination problem becomes more pronounced in a multi-cell environment and that realistic channel conditions are often correlated due to sparse scattering \cite{kim:twc:25}, there exists a need for a new pilot design approach that incorporates more generalized system environments.

This paper proposes a novel pilot design that simultaneously accounts for 1-bit massive MIMO, multi-cell environments, and spatially correlated channels. In such a system, quantization distortion and pilot contamination pose an interlocking performance bottleneck, making the joint optimization of both problems essential. To address this, we develop a novel framework that first transforms the intractable MSE into a tractable form via a low signal-to-noise ratio (SNR) approximation and subsequently solve the approximated problem by the FP approach. This constitutes the first practical solution for jointly mitigating quantization distortion and pilot contamination to the best of our knowledge. Numerical results demonstrate the superior performance achieved by the proposed method in channel estimation as opposed to the baseline approaches in general multi-cell and correlated channel environments.

\emph{Notation}: a, $\boldsymbol{a}$, and $\mathbf{A}$ are scalar, vector, matrix, respectively. The superscripts $(\cdot)^{\mathsf{T}}$, $(\cdot)^{\mathsf{H}}$, and $(\cdot)^{-1}$ represent the transpose, Hermitian, and inverse of a matrix, respectively. The operator $\mathrm{tr}(\cdot)$ denotes the trace of a matrix, and $\mathbf{I}_N$ is the $N \times N$ identity matrix. The notation $\|\cdot\|_2$ refers to the $\ell_2$-norm, and $\mathbb{E}[\cdot]$ denotes the expectation operator. We use $\otimes$, $\mathrm{sign}(\cdot)$, and $\mathrm{vec}(\cdot)$ to denote the Kronecker product, the sign function, and the vectorization operator, respectively.

\section{System Models}

\subsection{Signal Model}
We consider an uplink 1-bit massive MIMO system consisting of $L$ cells. Each cell $\ell \in \{1, \dots, L \}$ includes a base station (BS) equipped with $M$ antennas. Each antenna is paired with two 1-bit ADCs to separately process the real and imaginary parts of the received signal \cite{yongzhi:tsp:17}. Each cell includes $K$ users, each equipped with a single antenna. The full coherence bandwidth is reused across all users.
The channel for user $k \in \{1, \dots, K\}$ in cell $i$ to cell $\ell$, denoted as $\boldsymbol{h}_{\ell ik} \in \mathbb{C}^M$, is expressed as
\begin{align}
    \boldsymbol{h}_{\ell ik} = \sqrt{\beta_{\ell ik}}\boldsymbol{g}_{\ell ik}, \label{channel model}
\end{align}
where $\boldsymbol{g}_{\ell ik} \sim \mathcal{CN}(0, \mathbf{R}_{\ell ik})$ describes the small-scale fading, 
$\beta_{\ell ik}$ is the large-scale fading coefficient, and $\mathbf{R}_{\ell ik} \in \mathbb{C}^{M \times M}$ represents the covariance matrix. $\mathbf{R}_{\ell ik} \neq {\bf{I}}$ implies that each channel is inherently correlated with respect to the receiving antennas. 

During the training phase of each cell, all $K$ users simultaneously transmit their corresponding pilot sequences, each consisting of $\tau$ symbols, to the BS.
Focusing on cell $\ell$ without loss of generality, the received pilot signal at the BS is
\begin{align}
    \boldsymbol{Y}_{\ell} &= \sum_{k=1}^{K} \sqrt{\beta_{\ell \ell k}} \boldsymbol{g}_{\ell \ell k} \boldsymbol{\phi}_{\ell k}^\top + \underbrace{\sum_{i \neq \ell} \sum_{k=1}^{K} \sqrt{\beta_{\ell ik}} \boldsymbol{g}_{\ell ik}\boldsymbol{\phi}_{ik}^\top}_{\text{pilot interference}} + \boldsymbol{Z}_{\ell} \\
    &= \mathbf{H}_{\ell \ell} \mathbf{\Phi}_{\ell}^\top + \sum_{i \neq \ell} \mathbf{H}_{\ell i} \mathbf{\Phi}^\top + \boldsymbol{Z}_{\ell}, \label{receive signal}
\end{align}
where $\boldsymbol{\phi}_{\ell k} \in \mathbb{C}^\tau$ represents the pilot sequence vector of user $k$, $\mathbf{H}_{\ell \ell} = [\sqrt{\beta_{\ell \ell 1}}\boldsymbol{g}_{\ell \ell 1}, \dots, \sqrt{\beta_{\ell \ell K}}\boldsymbol{g}_{\ell \ell K}]$ denotes the channel matrix, and $\mathbf{\Phi}_{\ell}^\top = [\boldsymbol{\phi}_{\ell 1}^\top; \dots;\boldsymbol{\phi}_{\ell K}^\top]$ is the corresponding pilot matrix. $\boldsymbol{Z}_{\ell} \sim \mathcal{CN}(0, \sigma^2 \mathbf{I}_{\tau M})$ represents the additive white Gaussian noise (AWGN) matrix.
The pilot sequence $\boldsymbol{\phi}_{\ell k}$ is designed under the following constraints
\begin{align}
    \boldsymbol{\phi}_{\ell k} \in \mathbb{C}^\tau, \quad \| \boldsymbol{\phi}_{\ell k} \|^2_{2} \leq \tau P, \quad \forall \ell, \forall k. \label{power constraint}
\end{align}
where $P$ is the power constraint.

Next, vectorizing \eqref{receive signal}, the received signal \eqref{receive signal} is re-expressed as
\begin{align}
    \boldsymbol{y}_{\ell} &= \bar{\mathbf{\Phi}}_{\ell} \bar{\boldsymbol{h}}_{\ell \ell} + \sum_{i \neq \ell} \bar{\mathbf{\Phi}}_i \bar{\boldsymbol{h}}_{\ell i} + \boldsymbol{z}_{\ell}, \label{vetorized signal}
\end{align}
where $\boldsymbol{y}_{\ell} = \mathrm{vec}(\boldsymbol{Y}_{\ell})$, $\bar{\mathbf{\Phi}}_{\ell} = \mathbf{\Phi}_{\ell}\otimes \mathbf{I}_M$, $\bar{\boldsymbol{h}}_{\ell i} = \mathrm{vec}(\mathbf{H}_{\ell i})$, and $\boldsymbol{z}_{\ell} = \mathrm{vec}(\boldsymbol{Z}_{\ell})$.  
After passing through the 1-bit ADCs, the quantized received signal $\boldsymbol{b}_{\ell}$ is given by
\begin{align}
    \boldsymbol{b}_{\ell} = \mathcal{Q}(\boldsymbol{y}_{\ell}) = \mathcal{Q}\left(\bar{\mathbf{\Phi}}_{\ell} \bar{\boldsymbol{h}}_{\ell \ell} + \sum_{i \neq \ell} \bar{\mathbf{\Phi}}_i \bar{\boldsymbol{h}}_{\ell i} + \boldsymbol{z}_{\ell}\right). \label{quantized signal}
\end{align}
The 1-bit quantization operation in $\mathcal{Q}(\cdot)$ is defined as
$\mathcal{Q}(\cdot) = \frac{1}{\sqrt{2}} \left( \mathrm{sign}\left(\Re(\cdot)\right) + j \, \mathrm{sign}\left(\Im(\cdot)\right) \right)
$. As a result, output $\boldsymbol{b}_{\ell}$ is drawn from the set $\boldsymbol{b}_{\ell} \in \frac{1}{\sqrt{2}} \{1 + j, 1 - j, -1 + j, -1 - j \}.$

\subsection{Bussgang-Aided Channel Estimation}
The signal received through 1-bit ADCs is subject to a non-linear operation (specifically $\mathrm{sign}(\cdot)$ operation \eqref{quantized signal}).
This non-linearity complicates the subsequent analysis of estimation performance. 
To address these challenges, we exploit the fact that the received signal $\boldsymbol{y}_{\ell}$ follows a Gaussian distribution \cite{yongzhi:tsp:17}. 
By applying the Bussgang decomposition \cite{yongzhi:tsp:17, wan:twc:20}, the non-linear operation is transformed into a statistically equivalent linear model.
To be specific, $\boldsymbol{b}_{\ell}$ in \eqref{quantized signal} can be characterized as 
\begin{align}
    \boldsymbol{b}_{\ell} = \mathbf{A}_{\ell}\boldsymbol{y}_{\ell} + \boldsymbol{e}_{\ell}, \label{Bussgang decomposition}
\end{align}
where $\mathbf{A}_{\ell} \in \mathbb{C}^{\tau M \times \tau M}$ represents the linear operation matrix based on the LMMSE estimation, and $\boldsymbol{e}_{\ell} \in \mathbb{C}^{\tau M}$ denotes the quantization noise, which is uncorrelated with both $\boldsymbol{y}_{\ell}$ and $\boldsymbol{g}_{\ell}$ according to the Bussgang theorem. The autocorrelation matrix of $\boldsymbol{y}_{\ell}$ is obtained as
\begin{align}
    \mathbf{R}_{\boldsymbol{y}_{\ell}} &= \mathbb{E}[\boldsymbol{y}_{\ell} \boldsymbol{y}_{\ell}^\mathsf{H}] = \bar{\mathbf{\Phi}}_{\ell} \mathbf{R}_{\boldsymbol{h}_{\ell \ell}} \bar{\mathbf{\Phi}}_{\ell}^\mathsf{H} + \sum_{i \neq \ell} \bar{\mathbf{\Phi}}_i \mathbf{R}_{\boldsymbol{h}_{\ell i}} \bar{\mathbf{\Phi}}_i^\mathsf{H} + \sigma^2 \mathbf{I}_{\tau M}, \label{Ry}
\end{align}
where $\mathbf{R}_{\boldsymbol{h}_{\ell i}} = \mathrm{diag}\{\mathbf{R}_{\ell i1}, \mathbf{R}_{\ell i2}, \cdots, \mathbf{R}_{\ell iK}\}$ is the autocorrelation matrix of the channel coefficients for all the users to cell $\ell$.
The cross-correlation matrix between $\boldsymbol{y}_{\ell}$ and $\boldsymbol{b}_{\ell}$ is given by
\begin{align}
\mathbf{R}_{\boldsymbol{y}_{\ell}\boldsymbol{b}_{\ell}} = \mathbb{E}[\boldsymbol{y}_{\ell}\boldsymbol{b}_{\ell}^\mathsf{H}] = \sqrt{\frac{2}{\pi}} \mathbf{R}_{\boldsymbol{y}_{\ell}} \boldsymbol{\Sigma}_{\boldsymbol{y}_{\ell}}^{-\frac{1}{2}}, \label{Ryb}
\end{align}
where $\boldsymbol{\Sigma}_{\boldsymbol{y}_{\ell}}$ is the diagonal matrix of $\mathbf{R}_{\boldsymbol{y}_{\ell}}$. The linear operation matrix $\mathbf{A}_{\ell}$ can be written as
\begin{align}
    \mathbf{A}_{\ell} = \mathbf{R}^\mathsf{H}_{\boldsymbol{y}_{\ell} \boldsymbol{b}_{\ell}} \mathbf{R}_{\boldsymbol{y}_{\ell}}^{-1} = \sqrt{\frac{2}{\pi}} \boldsymbol{\Sigma}_{\boldsymbol{y}_{\ell}}^{-1/2}. \label{A}
\end{align}
By substituting \eqref{A} into \eqref{Bussgang decomposition}, we arrive at
\begin{align}
    \boldsymbol{b}_{\ell} &= \sqrt{\frac{2}{\pi}} \boldsymbol{\Sigma}_{\boldsymbol{y}_{\ell}}^{-1/2} 
    \left( \bar{\mathbf{\Phi}}_{\ell} \bar{\boldsymbol{h}}_{\ell \ell} + \sum_{i \neq \ell} \bar{\mathbf{\Phi}}_i \bar{\boldsymbol{h}}_{\ell i} \right) + \bar{\boldsymbol{z}}_{\ell}, \label{b}
\end{align}
where $\bar{\boldsymbol{z}}_{\ell} = \sqrt{\frac{2}{\pi}} \boldsymbol{\Sigma}_{\boldsymbol{y}_{\ell}}^{-1/2} \boldsymbol{z}_{\ell} + \boldsymbol{e}_{\ell}$ is the effective noise and is also uncorrelated with both $\bar{\boldsymbol{h}}_{\ell i}$ and $\bar{\boldsymbol{h}}_{\ell \ell}$.

We adopt the BLMMSE estimator \cite{yongzhi:tsp:17} to estimate the channel from $\boldsymbol{b}_{\ell}$. 
BLMMSE is known as a computationally efficient solution for channel estimation under the Bussgang linearized system model, while also ensuring robust performance \cite{yongzhi:tsp:17, wan:twc:20}. 
BLMMSE computes
\begin{align}
    \hat{\boldsymbol{h}}_{\ell \ell k} = \mathbf{R}_{\boldsymbol{b}_{\ell}\boldsymbol{h}_{\ell \ell k}}\mathbf{R}_{\boldsymbol{b}_{\ell}}^{-1}\boldsymbol{b}_{\ell}, \label{h_hat}
\end{align}
where
\begin{align}
    \mathbf{R}_{\boldsymbol{b}_{\ell}\boldsymbol{h}_{\ell \ell k}}
    &= \sqrt{\frac{2}{\pi}}\beta_{\ell \ell k} \left( \boldsymbol{\phi}_{\ell k}^\mathsf{H} \otimes \mathbf{R}_{\ell \ell k}\right)\boldsymbol{\Sigma}_{\boldsymbol{y}_{\ell}}^{-\frac{1}{2}}. \label{Rbh}
\end{align}
The correlation matrix of the quantized signal $\boldsymbol{b}_{\ell}$, denoted as $\mathbf{R}_{\boldsymbol{b}_{\ell}}$, is obtained by applying the arcsine law as
\begin{align}
&\mathbf{R}_{\boldsymbol{b}_{\ell}} = \frac{2}{\pi} \Bigg[ \arcsin\left(\boldsymbol{\Sigma}_{\boldsymbol{y}_{\ell}}^{-\frac{1}{2}} \Re(\mathbf{R}_{\boldsymbol{y}_{\ell}}) \boldsymbol{\Sigma}_{\boldsymbol{y}_{\ell}}^{-\frac{1}{2}} \right) \notag\\& \quad\quad +j \, \arcsin\left(\boldsymbol{\Sigma}_{\boldsymbol{y}_{\ell}}^{-\frac{1}{2}} \Im(\mathbf{R}_{\boldsymbol{y}_{\ell}}) \boldsymbol{\Sigma}_{\boldsymbol{y}_{\ell}}^{-\frac{1}{2}} \right) \Bigg].  \label{Rb}
\end{align}
Accordingly, the sum of MSEs across all the uplink users is characterized as
\begin{align}
    \text{MSE}_{\Sigma} 
    = \sum_{(\ell,k)}{\beta_{\ell \ell k}\text{tr}\left(\mathbf{R}_{\ell \ell k}\right)}-\sum_{(\ell,k)}{\text{tr}\left(\mathbf{R}_{\boldsymbol{b}_{\ell}\boldsymbol{h}_{\ell \ell k}}\mathbf{R}_{\boldsymbol{b}_{\ell}}^{-1}\mathbf{R}_{\boldsymbol{b}_{\ell}\boldsymbol{h}_{\ell \ell k}}^\mathsf{H}\right)}. \label{MSE}
\end{align}

\subsection{Pilot Design Problem Formulation}
We aim to minimize $\text{MSE}_{\Sigma}$ in \eqref{MSE} by coordinating the pilot sequences $\boldsymbol{\phi}_{\ell k}$,  $\forall \ell, k$. 
We observe from \eqref{MSE} that $\mathbf{R}_{\boldsymbol{b}_{\ell}\boldsymbol{h}_{\ell \ell k}}$ and $\mathbf{R}_{\boldsymbol{b}_{\ell}}$ are functions related to $\{\boldsymbol{\phi}\} = \{ \boldsymbol{\phi}_{11}, \cdots ,\boldsymbol{\phi}_{LK} \}$, so the pilot design problem is formulated to 
\begin{subequations}
\begin{align}
& \underset{\{{\boldsymbol{\phi}}\}}{\text{maximize}} \quad \sum_{(\ell,k)}{\text{tr}\left(\mathbf{R}_{\boldsymbol{b}_{\ell}\boldsymbol{h}_{\ell \ell k}}\mathbf{R}_{\boldsymbol{b}_{\ell}}^{-1}\mathbf{R}_{\boldsymbol{b}_{\ell}\boldsymbol{h}_{\ell \ell k}}^\mathsf{H}\right)} \label{objective function}
\\& \text{subject to}  \quad \|\boldsymbol{\phi}_{\ell k}\|_2^2 \leq \tau P, \quad \forall (\ell,k). \label{constraint}
\end{align}
\end{subequations}
However, this optimization problem is non-convex and in an intractable form, due to i) the presence of the element-wise arcsine function in $\mathbf{R}_{\boldsymbol{b}_{\ell}}$ and ii) the multiple matrix ratio form. To address these, we reformulate \eqref{objective function} into a more tractable form in the next section.
\section{Problem Reformulation}

\subsection{Low SNR Approximation}
By considering that the performance degradation caused by 1-bit ADCs is alleviated in the low SNR regime (with only a $1.96$ dB loss as SNR approaches $0$ \cite{mezghani:isit:09}), it is suitable to operate the 1-bit massive MIMO system in the low SNR regime.
Motivated by this, we derive an approximate MSE expression under the assumption of the low SNR regime. 
In Section V, we demonstrate that, despite the low SNR approximation, our pilot design method performs effectively even in the high SNR regime.
By applying the first-order Taylor approximation of the arcsine function near $0$, we have $\arcsin(x) \approx x$. Upon this, we approximate $\mathbf{R}_{\boldsymbol{b}_{\ell}}$ as
\begin{equation}
\begin{aligned}
\mathbf{R}_{\boldsymbol{b}_{\ell}} 
& \approx \frac{2}{\pi}
\boldsymbol{\Sigma}_{\boldsymbol{y}_{\ell}}^{-\frac{1}{2}} 
\mathbf{R}_{\boldsymbol{y}_{\ell}} 
\boldsymbol{\Sigma}_{\boldsymbol{y}_{\ell}}^{-\frac{1}{2}} 
+ \left(1 - \frac{2}{\pi} \right) \mathbf{I}_{\tau M}.
\end{aligned}
\end{equation}
By using the approximation result of $\mathbf{R}_{\boldsymbol{b}_{\ell}}$, we substitute it into \eqref{objective function}, allowing us to reformulate the optimization problem as
\begin{subequations} \label{Re: full_problem}
\begin{align}
& \underset{\{{\boldsymbol{\phi}}\}}{\text{maximize}} \quad \sum_{(\ell,k)}{\text{tr}\left(\mathbf{A}_{\ell k}\mathbf{B}_{\ell}^{-1}\mathbf{A}^\mathsf{H}_{\ell k}\right)} \label{Re: objective function}
\\& \text{subject to}  \quad \|\boldsymbol{\phi}_{\ell k}\|_2^2 \leq \tau P, \quad \forall (\ell,k), \label{Re: constraint}
\end{align}
\end{subequations}
where $\mathbf{A}_{\ell k} \in \mathbb{C}^{M\times \tau M}$ and $\mathbf{B}_{\ell} \in \mathbb{C}^{\tau M\times \tau M} $ are given by
\begin{align}
& \mathbf{A}_{\ell k} = \beta_{\ell \ell k} \left( \boldsymbol{\phi}_{\ell k}^\mathsf{H} \otimes \mathbf{R}_{\ell \ell k}\right), \; \mathbf{B}_{\ell} =  \mathbf{R}_{\boldsymbol{y}_{\ell}} + \left( \frac{\pi}{2} -1\right) \boldsymbol{\Sigma}_{\boldsymbol{y}_{\ell}}. 
\end{align}
Subsequently, we transform \eqref{Re: objective function}, so as to make the multiple-ratio optimization problem tractable. It turns out that the proposed approximation also works for the high-SNR regime, as discussed later in Section V.

\subsection{Matrix Quadratic Transform}
The objective function in \eqref{Re: full_problem} admits a multiple-ratio fractional form, which is difficult to deal with directly. To address this, we exploit the quadratic transform technique, which has been recently introduced in the matrix FP framework \cite{kaiming_coordinated, kaiming_matrix_FP}. 
This enables us to equivalently reformulate the problem by simultaneously decoupling the numerators and denominators of the multiple ratios. 
Applying this, \eqref{Re: full_problem} is equivalently transformed to
\begin{subequations} \label{final problem}
\begin{align}
&\underset{\{{\boldsymbol{\phi}}\}, \{\mathbf{\Lambda}\}}{\text{maximize}} 
&& \sum_{(\ell,k)} \text{tr} \left(2 \Re \left\{\mathbf{A}_{\ell k} \mathbf{\Lambda}_{\ell k}\right\} - \mathbf{\Lambda}_{\ell k}^\mathsf{H} \mathbf{B}_{\ell} \mathbf{\Lambda}_{\ell k}\right) \label{eq: transform}\\
&\text{subject to} 
&& \|\boldsymbol{\phi}_{\ell k}\|_2^2 \leq \tau P, \label{eq: transformed_power constraint}\\
&&& \mathbf{\Lambda}_{\ell k} \in \mathbb{C}^{\tau M \times M}, \quad \forall (\ell,k),
\end{align}
\end{subequations}
where $\mathbf{\Lambda}_{\ell k}$ is defined as an auxiliary variable for $\mathbf{A}_{\ell k}\mathbf{B}_{\ell}^{-1}\mathbf{A}^\mathsf{H}_{\ell k}$. 
The equivalence between \eqref{Re: objective function} and \eqref{eq: transform} is discussed in detail in \cite{kaiming_matrix_FP}. Also, it can be established by substituting the optimal $\{\mathbf{\Lambda}\} = \{ \mathbf{\Lambda}_{11}, \cdots , \mathbf{\Lambda}_{LK} \}$ while fixing $\{\boldsymbol{\phi}\}$.

\section{Pilot Design}

\subsection{Pilot Design Algorithm}

To solve the optimization problem in \eqref{eq: transform}, we employ an alternating optimization approach that iteratively optimizes $\{\boldsymbol{\phi}\}$ and $\{\mathbf{\Lambda}\}$. We now describe the update procedure for each variable. When $\{\boldsymbol{\phi}\}$ is fixed, the optimal $\{\mathbf{\Lambda}\}$ is obtained by differentiating the objective function with respect to $\mathbf{\Lambda}_{\ell k}$, resulting in $\mathbf{\Lambda}^*_{\ell k} = \mathbf{B}_{\ell}^{-1} \mathbf{A}^\mathsf{H}_{\ell k}$.
Thereafter, we update $\{\boldsymbol{\phi}\}$ while keeping $\{\mathbf{\Lambda}\}$ fixed. Before this, we rewrite the objective function to allow the optimal value of $\{\boldsymbol{\phi}\}$ to be derived in closed form.
Specifically, the first terms of the objective function in \eqref{final problem} is expressed as 
\begin{align}
\sum_{(\ell,k)} \text{tr} \left( 2\Re \left\{ \mathbf{A}_{\ell k} \mathbf{\Lambda}_{\ell k} \right\} \right)
= \sum_{(\ell,k)} 2\Re \left\{ \boldsymbol{\phi}_{\ell k}^{\mathsf{H}} \boldsymbol{v}_{\ell k} \right\}, \label{first term of objective function}
\end{align}
where the $(p,q)$th entry of $\boldsymbol{v}_{\ell k} \in \mathbb{C}^\tau$ is expressed as
\begin{align}
&\boldsymbol{v}_{\ell k}[p,q] \notag\\
&=\beta_{\ell \ell k}\sum^M_{m=1} \mathbf{R}_{\ell \ell k}[m,:]\mathbf{\Lambda}_{\ell k}[(p-1)M+1:iM,m+(q-1)M].
\end{align}

Next, the second terms of the objective function in \eqref{final problem} is also expressed as 
\begin{align}
& \sum_{(\ell,k)} \text{tr} \left( \mathbf{\Lambda}_{\ell k}^{\mathsf{H}} \mathbf{B}_{\ell} \mathbf{\Lambda}_{\ell k} \right)
= \sum_{(\ell,k)} \boldsymbol{\phi}_{\ell k}^{\mathsf{H}} \mathbf{M}_{\ell k} \boldsymbol{\phi}_{\ell k} +  \text{const}. \label{second term of objective function}
\end{align}
Here, we have
\begin{align}
& \widetilde{\mathbf{\Lambda}}_{\ell k} = {\mathbf{\Lambda}}_{\ell k}{\mathbf{\Lambda}}_{\ell k}^\mathsf{H}, \\
& \mathbf{S}_{\ell ijk}[p,q] \notag\\&= \beta_{\ell ij}\sum^M_{m=1}  \mathbf{R}_{\ell ij}[m,:]\widetilde{\mathbf{\Lambda}}_{\ell k}[(p-1)M+1:pM,m+(q-1)M],\\
& \mathbf{T}_{\ell ijk}[p,q] \notag \\
& =
\begin{cases}
\beta_{\ell ij}\sum^M_{m=1}  \mathbf{R}_{\ell ij}[m,m] \\ \quad\quad\quad \times \widetilde{\mathbf{\Lambda}}_{\ell k}[m+(p-1)M,m+(q-1)M] , & \text{if } p = q,\\
0, & \text{otherwise},
\end{cases}\\
& \mathbf{M}_{\ell k} = \sum_{(i,j)} \left(  \mathbf{S}_{\ell ijk} + \left( \frac{\pi}{2} -1\right)\mathbf{T}_{\ell ijk}\right).
\end{align}
Finally, we note that $\text{const} = \sigma^2\sum_{(\ell, k))} \text{tr}\left( \widetilde{\mathbf{\Lambda}}_{\ell k}\right)$ does not depend on $\{\boldsymbol{\phi}\}$.
Leveraging the derived form, we obtain the optimal pilot $\boldsymbol{\phi}^*_{\ell k}$ as $\boldsymbol{\phi}_{\ell k}^* = \left( \mathbf{M}_{\ell k} + \eta_{\ell k} \mathbf{I}_{\tau} \right)^{-1} \boldsymbol{v}_{\ell k}$.
Here, the Lagrangian multiplier $\eta_{\ell k}$ is adjusted by using the bisection search method to ensure that $\boldsymbol{\phi}_{\ell k}^*$ does not exceed the power constraint \eqref{eq: transformed_power constraint}. By iterating the optimization process until the MSE converges, we find the local optimum $\boldsymbol{\phi}^*_{\ell k}$. 

\subsection{Convergence and Complexity Analysis}

The proposed alternating optimization algorithm can be interpreted within the Minorization-Maximization (MM) framework \cite{mm_algorithm}. This ensures that the objective function is monotonically non-decreasing in each step, guaranteeing convergence to a stationary point of the problem.

The computational complexity is dominated by the matrix inversions required to update the auxiliary variable $\boldsymbol{\Lambda^*}$ and the pilot sequence $\boldsymbol{\phi^*}$. The overall computational complexity of pilot design is given by $\mathcal{O}(K^2L^2\tau^2M^2 + KL\tau^3M^3)$. Notably, our pilot design is based on the channel covariance matrix, which changes on a slow time scale and thus does not need to be computed instantaneously. This inherent temporal stability allows the pilot design to be performed offline or infrequently, resulting in an overall computational complexity that remains low.

\section{Numerical Results}
We evaluate the proposed Bussgang-aided FP (BFP) pilot design in a $L=7$-cell hexagonal layout with 0.5\,km inter-site distance, where no user is allowed to be closer than 35\,m to any BS. Each BS has $M=64$ antennas and each cell serves $K=4$ single-antenna users uniformly at random. The pilot length and user transmit power are set to $\tau=10$ and $P=23$\,dBm, respectively. The noise power spectral density and bandwidth are $-169$\,dBm/Hz and 20\,MHz. $\beta_{\ell ik}$ (\text{dB}) is modeled as $\beta_{\ell ik} = 128.1 + 37.6 \log_{10} \left( d_{\ell ik} \right) + \psi_{\ell ik}$ \cite{yu_pathloss}.
Here, $d_{\ell ik}$ (\text{km}) represents the distance between the BS and the corresponding user, and $\psi_{\ell ik}$ represents shadow fading and follows log-normal distribution parameters $\mathcal{N}(0, 64)$. In addition, the correlated channel is modeled using an exponential model \cite{loyka_exponential_matrix}, where the channel covariance matrix $\mathbf{R}_{\ell ik}$ is expressed as
\begin{align}
\mathbf{R}_{\ell ik}[{m,n}] =
\begin{cases} 
    \omega_{\ell ik}^{m-n}, & \text{if } m \geq n; \\
    \mathbf{R}^\mathsf{H}_{\ell ij}[{m,n}], & \text{otherwise},
\end{cases}
\end{align}
where $ \omega = \nu e^{j\theta}$, $\nu = 0.5$, and $\theta$ follows an i.i.d uniform distribution $U[0, 2\pi)$. The channel estimation performance was evaluated based on the normalized MSE, defined as
\begin{align}
    \text{NMSE} = \frac{\sum_\ell \sum_k \| \boldsymbol{h}_{\ell \ell k} - \hat{\boldsymbol{h}}_{\ell \ell k } \|_2^2}{\sum_\ell \sum_k \| \boldsymbol{h}_{\ell \ell k} \|_2^2}.
\end{align}

For the baseline methods, we consider the followings. 
i) the discrete Fourier transform (DFT) pilots, ii) the random pilots, iii) the FP based pilots \cite{kaiming_coordinated}, iv) the stochastic variance reduced gradient projection (SVGRP) pilots \cite{wu:icassp:20}, and v) the NMSE lower bound \cite{yi_one_bit_Rayleigh}. 
To be specific, in the DFT pilots, we constructed a $KL$ DFT matrix and assigned pilot sequences by selecting rows with spacing $L$. In the FP based pilots \cite{kaiming_coordinated}, we directly apply the FP algorithm to 1-bit ADC systems without any modifications.
For the channel estimation, we adopt the BLMMSE estimator \cite{yongzhi:tsp:17}.  

\begin{figure}[!t]
    \centerline{\resizebox{1\columnwidth}{!}{\includegraphics{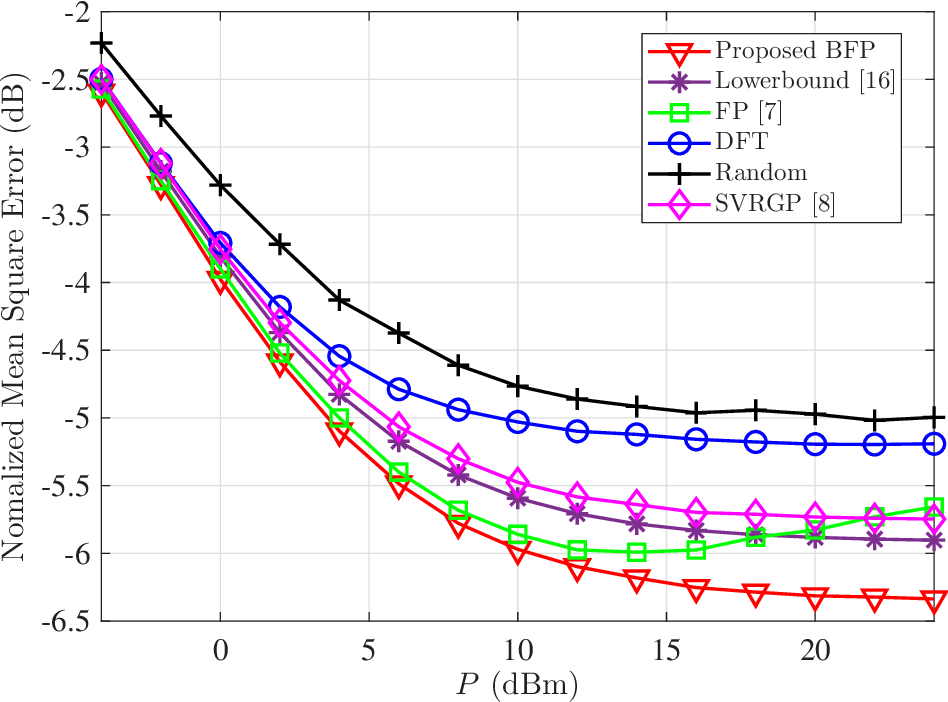}}}
    \caption{NMSE vs. transmit power}
    \label{NMSE vs. power} 
\end{figure}

In Fig.~\ref{NMSE vs. power}, we illustrate the NMSE vs. the transmit power. 
Since the NMSE lower bound was derived under a single-cell scenario with max-min power control, to evaluate this in our environment, we adapt our setup to be fitted to \cite{yi_one_bit_Rayleigh}.
Specifically, focusing on a certain cell, we forcefully fix $\beta_{\ell ik}$ values to the same constant and assume the uncorrelated Rayleigh fading, i.e., $\mathbf{R}_{\ell ik} = {\bf{I}}$. 
Additionally, we treat interference coming from the other cells as additive Gaussian noise, which not only simplifies the analysis but also effectively extends the existing NMSE lower bound to the multi-cell scenario.
As observed in Fig.~\ref{NMSE vs. power}, the proposed BFP method outperforms other methods as well as the NMSE lower bound across all transmit power levels in terms of NMSE.
Since this lower bound was obtained under the single-cell and uncorrelated case, the pilot contamination in a multi-cell and the impact on channel correlation was not incorporated.
On the contrary to that, our BFP method is able to alleviate the inter-cell interference and adaptively capture the correlation across channels. Although the FP and SVGRP method also considers these aspects, it lacks the capability to account for the autocorrelation in the modified MSE caused by 1-bit ADCs. Therefore, our method achieves superior performance as it is designed to be well-suited for this environment.

It is noteworthy that our design, derived from a low SNR approximation, remains effective even in the high SNR regime. The key reason is that the validity of the approximation depends on both SNR and the spatial correlation of the channel, with the latter being the dominant factor. In practical environments, as described by the exponential correlation model, the correlation between antennas typically decays rapidly, which ensures that our approximation remains accurate. As a result, the proposed optimization not only accounts for the 1-bit quantization distortion neglected by baselines but also preserves inter-cell interference, the dominant limiting factor at high SNR. This leads to more accurate channel estimation, particularly where quantization error rather than AWGN becomes the main impairment. These results align with Fig.~\ref{NMSE vs. power}.

\begin{figure}[!t]
     \centerline{\resizebox{1\columnwidth}{!}{\includegraphics{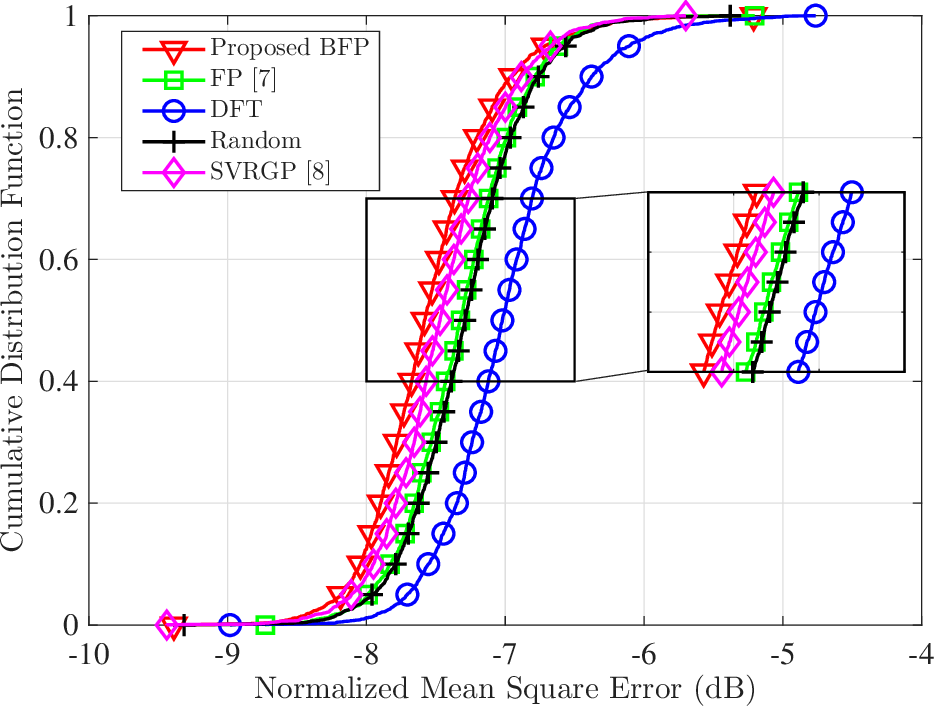}}}
    \caption{CDF of NMSE}
    \label{NMSE vs. CDF} 
\end{figure}

For more comprehensive performance evaluation, we also plot the cumulative distribution function (CDF) of the NMSE in Fig.~\ref{NMSE vs. CDF}. In this case, we conduct a system-level simulation, wherein we use general $\beta_{\ell ik}$ following \cite{yu_pathloss}. The NMSE lower bound is not plotted as it is incompatible with the considered setup. 
Fig.~\ref{NMSE vs. CDF} also shows that the proposed BFP method achieves the best NMSE over the other baseline methods.

\section{Conclusion}
In this paper, we proposed a coordinated pilot design for 1‑bit massive MIMO with multi‑cell correlated channels. Using a low‑SNR approximate MSE and a matrix FP reformulation, we obtained closed‑form updates enabling practical implementation. Simulations in realistic cellular layouts showed consistent NMSE gains over baseline pilots across SNRs. Our framework can be extended to any network topology, such as the cell-free setting. Thus, future work can include an analysis of association mechanisms and their joint optimization with pilot design.

\bibliographystyle{IEEEtran}
\bibliography{ref_1bit_pilot}

\end{document}